\documentclass{article}
\usepackage{spconf,amsmath,graphicx}

\usepackage{enumitem}
\setlist{nosep, leftmargin=14pt}
\usepackage{scrextend}
\usepackage{hyperref}
\usepackage{amsfonts}
\usepackage{booktabs}
\usepackage{siunitx}

\title{Automated triage of COVID-19 from various lung abnormalities \\ using chest CT features}

\name{Dor Amran$^{\star}$, Maayan Frid-Adar$^{\star}$, Nimrod Sagie$^{\star \dagger}$, Jannette Nassar$^{\star}$, Asher Kabakovitch$^{\star}$, Hayit Greenspan$^{\star \dagger}$}

\address{$^{\star}$ RADLogics Inc., Boston, MA\\  
         $^{\dagger}$ Department of Biomedical Engineering, Tel-Aviv University, Tel-Aviv, Israel}
\begin{document}
\maketitle
\begin{abstract}
The outbreak of COVID-19 has lead to a global effort to decelerate the pandemic spread. For this purpose chest computed-tomography (CT) based screening and diagnosis of COVID-19 suspected patients is utilized, either as a support or replacement to reverse transcription–polymerase chain reaction (RT-PCR) test. In this paper, we propose a fully automated AI based system that takes as input chest CT scans and triages COVID-19 cases. More specifically, we produce multiple descriptive features, including lung and infections statistics, texture, shape and location, to train a machine learning based classifier that distinguishes between COVID-19 and other lung abnormalities (including community acquired pneumonia). We evaluated our system on a dataset of 2191 CT cases and demonstrated a robust solution with 90.8\% sensitivity at 85.4\% specificity with 94.0\% ROC-AUC. In addition, we present an elaborated feature analysis and ablation study to explore the importance of each feature. 
\end{abstract}

\begin{figure}[t!]
 \centerline{\includegraphics[width=8.5cm]{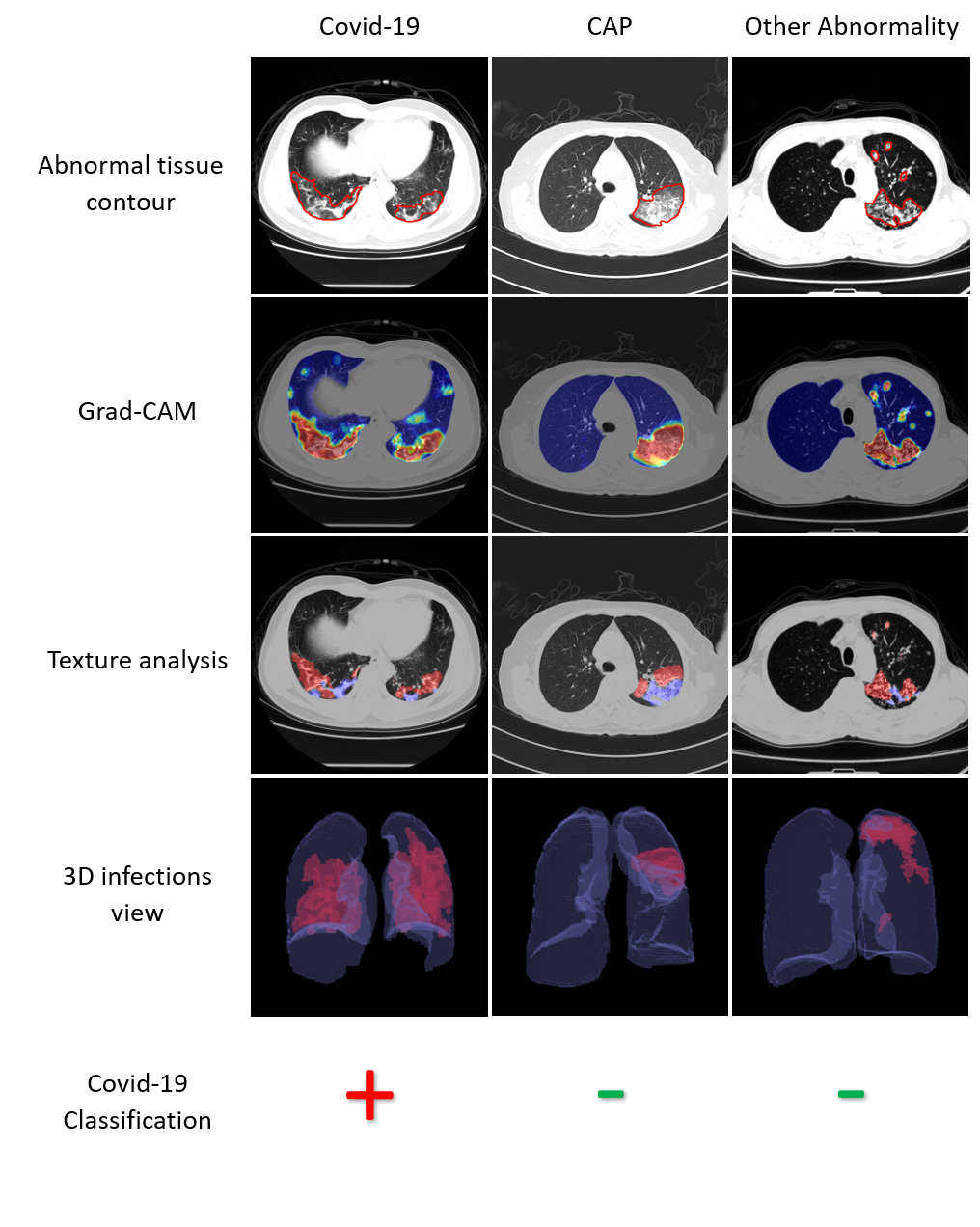}}
  \caption{System segmentation output maps (rows 1-4) and final classification decision (row 5). Three cases are shown with varying pathologies (columns). Visual differences include laterality, peripherality, volume and location of the detections.}
 \label{keyimages}
\end{figure}

\begin{figure*}
  \includegraphics[width=\textwidth]{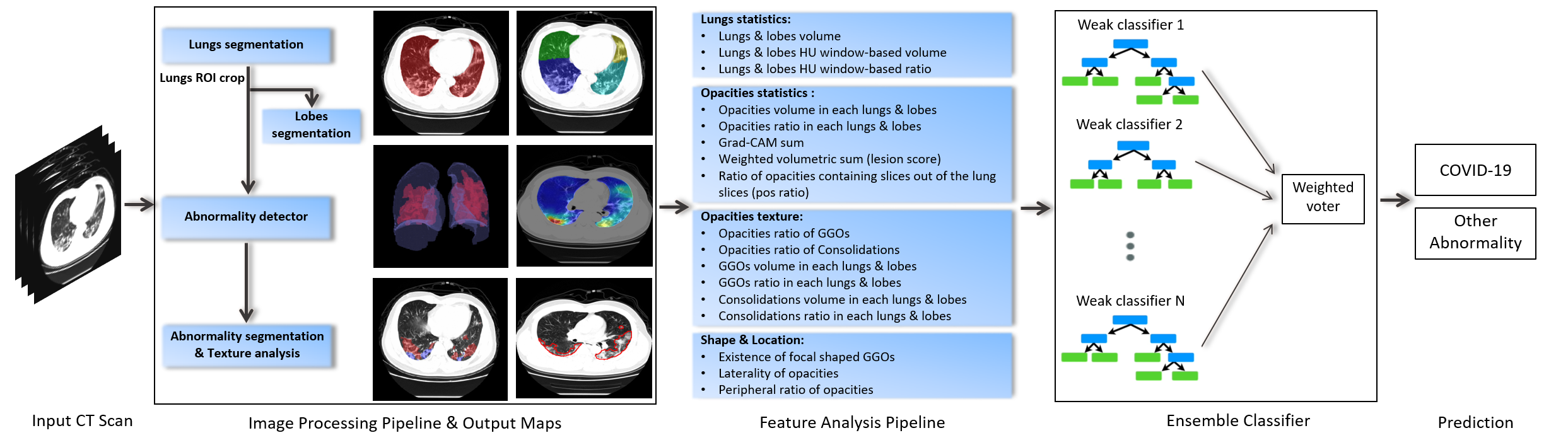}
  \caption{Chest CT scan classification process, depicting the image processing pipeline and output maps followed by the feature analysis pipeline and classification, in which this paper is focused.}
  \label{classsifierpipeline_boxes}
\end{figure*}

\section{Introduction}
\label{sec:intro}
COVID-19 pandemic is spreading worldwide, infecting millions of people and affecting everyday lives. In a majority of infected countries, virus containment steps include isolation of the infected population, thus, breaking the chain of infection. To do so, accurate, quick and efficient diagnosis and triage solutions are required to distinguish COVID-19 from both non-infected individuals and those who suffer other lung abnormalities. 
Chest CT imaging becomes an effective tool for that purpose as it produces results that are clinically actionable either for establishing a diagnosis or for guiding patient management, triage, or therapy.
Moreover, false-negative RT-PCR tests have been reported in patients with CT findings of COVID-19 who eventually tested positive after serial sampling \cite{Rubin2020}. Chest CT explicit clinical features are harnessed by physicians worldwide to interpret lung disease manifestations and produce a distinction between COVID-19 and other lung abnormalities \cite{Caruso2020, Wang2020}. 
Several studies have already been published since the pandemic outbreak utilizing artificial intelligence (AI) analysis of CT scans for diagnosis and quantification of COVID-19 \cite{shi2020review, Ozsahin2020}. Most studies dealing with screening of COVID-19 patients use a single Convolutional Neural Network (CNN) architecture for classification or segmentation to conclude whether an input case is COVID-19 or other (usually normal/other pneumonia type). Shi et al.\cite{Shi2020} argued that the classification decision of a single CNN is not robust enough to capture the variability of the lung disease manifestations, and suggested to learn clinically meaningful features as the basis to a machine learning classifier that distinguishes between COVID-19 and community acquired pneumonia (CAP). In this paper we present our classification system to discriminate between COVID-19 and various pathological conditions, including CAP. Figure \ref{keyimages} shows examples of system output maps (rows 1-4). Using these maps we produce a set of clinical features from which we generate the final classification (row 5), which is the focus of this current work. 
We use a multi-source large-scale dataset that holds cases containing various opacity manifested lung abnormalities gathered specifically to target this complex differentiation problem. We demonstrate the classification performance and analyze the clinical features contribution using both feature ablation experiments and feature distribution analysis. 

\section{Methodology}
\label{sec:format}
\subsection{Dataset Preparation}
Several public and private chest CT datasets were used, originating from various sources worldwide and numerous CT scanners to enable robustness 
(see Table \ref{datatable}). Input scans slice thickness ranged from 1mm to 10mm
with 0.5mm to 1.3mm 
pixel spacing. Each scan underwent alignment to "RAI+" orientation, clipping between $[-1000,0]$ Hounsfield Units (HU) and intensity normalization to $[0,1]$. Our dataset includes 2191 chest CT scans in total, divided into 1268 COVID-19 \& 923 non-COVID-19 lung abnormalities. 

\begin{table*}[ht!]
\centering
\resizebox{\linewidth}{!}{
 \begin{tabular}{llll} 
 \hline
 \textbf{Origin} & \textbf{Source} & \textbf{Subjects} & \textbf{Comments}  \\ 
 \hline
 China & CC-CCII \cite{Zhang2020}  & 772 COVID-19, 740 CAP & slice  thickness ${1-10}mm$\\ 
 \hline
 China & Private & 224 COVID-19, 5 lung abnormalities & slice thickness ${1,1.5,5}mm$ at one or more time points (between 1 and 5) \\ 
 \hline
US & Private & 68 lung abnormalities & slice thickness ${1-8} mm$, gathered before the COVID-19 pandemic  \\ 
 \hline
China & Private & 110 lung abnormalities & slice thickness ${5-10} mm$, gathered before the COVID-19 pandemic\\ 
\hline
Russia & Private & 272 COVID-19 & slice thickness ${1-5} mm$ \\ 
 \hline  \hline 
 \textbf{Total} & \multicolumn{2}{l}{\textbf{2191 chest CT scans: 1268 COVID-19 \& 923 non-COVID-19 lung abnormalities}} \\
\hline
\end{tabular}}
\caption{\label{datatable} Classifier datasets description.}
\end{table*}

\subsection{Image processing pipeline}
\label{ssec:subhead}
Figure \ref{classsifierpipeline_boxes} shows our system, which is composed of two main steps: an image processing pipeline and a feature analysis pipeline, which  produces the final classification.
The image processing pipeline takes as input a non-contrast chest ct scan and produces several segmentation output maps. It is comprised of several components: First, a lung segmentation step is performed using a 2-D U-Net to produce a 3-D Region of Interest (ROI) of the lungs. Second, a 3-D U-Net is employed on this ROI to provide lung lobes segmentation. Following these steps, abnormalities are detected using a weakly supervised CNN model as described in \cite{Gozes2020, Gozesmiccai2020}.
Furthermore, since bilateral and peripheral lung distribution ground-glass opacity (GGO) and consolidative opacities are the most common characteristics of COVID-19 \cite{Bernheim2020}, a 2-D U-Net++ was trained to provide a segmentation map that includes these categories.

\subsection{Feature analysis pipeline}
\label{Features_description}
In the feature analysis pipeline, we use the output segmentation maps to 
extract quantitative and descriptive clinical features. These features are then used to distinguish between COVID-19 and other lung abnormalities. We next describe the features extracted:\\
\textbf{Lungs statistics features:} volumetric lung features and HU analysis features. The volumetric features (calculated in $[cm^3]$) include the combined volume of both lungs, separate volume of the left/right lungs and the volume of each of the 5 lung lobes (henceforth called "anatomical structures"). HU changes could amount from a possible presence of lung opacity and should be therefore tracked. Analysis features include both the HU volume and ratio (calculated in $[\%]$) in division to the anatomical structures, detected in three HU windows: (a) low $[-1000,-950]$ (b) functional $[-950,-600]$ and (c) high $[-600,-250]$.

\textbf{Opacities statistics features:} volumetric ($[cm^3]$) and ratio ($[\%]$) features quantifying all detected abnormalities simultaneously in division to the anatomical structures.
Additional features are the "pos-ratio", i.e the ratio of abnormalities containing slices out of all lung slices, and Grad-CAM based features \cite{Gozes2020}. Grad-CAM features are extracted from the "Lung Abnormality detector" model activation maps and include the maps sum and volume weighted CNN model activations as a method of incorporating weakly supervised based abnormalities as a part of the source data.

\textbf{Opacities texture features:} abnormal texture detected is further analyzed and divided into GGOs and Consolidations.
For each of these opacities and anatomical structures the opacity volume ($[cm^3]$) and ratio ($[\%]$) are extracted. Furthermore, to determine the dominance of the opacity, it's fraction out of the entire abnormal volume is calculated.

\textbf{Shape \& Location features:} shape assessment is composed of a binary decision whether focal shaped GGOs ($<3 cm$ maximum axial diameter and of rounded morphology) were detected as a part of the abnormal tissue map. The location features include the abnormal tissue detected laterality classification  ($\in\{$unilateral\ left, unilateral\ right, bilateral\ $\}$) and peripheral ratio ($[\%]$). The first is a threshold based decision, in which we assert whether the abnormal tissue exceeds a pre-defined volume in a single (left/right) lung / both lungs. The second is determined by calculating the ratio of peripheral opacities (i.e overlapping with the dilated edge of the lungs inner surface, excluding the bronchial tree adjacent area) out of the entire abnormal volume detected.

\subsection{COVID-19 vs other abnormalities triage}
\label{ssec:subhead}
As shown in Figure \ref{classsifierpipeline_boxes}, once the image pipeline outputs and feature extraction steps are performed for all chest CT scans, a feature-based dataset is used to train a machine learning classifier to differentiate between COVID-19 and other lung abnormalities. We experiment with both a Random-Forest (RF) classifier and an AdaBoost classifier with several types of kernels; decision tree (DT), Linear, Polynomial and a radial basis function (RBF). During the training process an ensemble classifier is fitted. In each training iteration a weak classifier is added, fitting to the weighted train data. Weight is determined by previous weak classifiers performance on the same data, thus minimizing the misclassification error rate. The final ensemble model is composed of a linear combination of the aforementioned weak classifiers producing the COVID-19 probability. The classifier operating point probability threshold is then chosen for the final prediction, balancing the classifier's sensitivity and specificity metrics.
To optimize the training hyper-parameters, a gross to fine grid search is performed to select the best suited ensemble model, weak classifiers design, number of weak classifiers composing the ensemble, model learning rate and minimum number of cases per weak classifier split. Following the hyper-paramater optimization process the parameters ensuring the highest mean ROC-AUC and sensitivity balance were chosen.

\begin{center}
\begin{table*}[t]
\resizebox{\linewidth}{!}{%
\begin{tabular}{clccccccc} \toprule

    $Nº$ & $Classifier$                        &  $Sensitivity$         &  $Specificity$        &   $Precision$          &      $F1$             &   $Accuracy$          &      $AUC$             \\ \midrule
    1 & RF                                     & $0.822\pm0.015$        & $0.782\pm0.033$       & $0.837\pm0.025$        & $0.830\pm0.017$       & $0.778\pm0.014$       & $0.872\pm0.015$        \\
    2 & AdaBoost - SVM - Linear                & $0.781\pm0.045$        & $0.705\pm0.066$       & $0.786\pm0.030$        & $0.782\pm0.015$       & $0.740\pm0.017$       & $0.800\pm0.013$        \\
    3 & AdaBoost - SVM - Polynomial                  & $0.766\pm0.040$        & $0.760\pm0.084$       & $0.818\pm0.049$        & $0.789\pm0.006$       & $0.602\pm0.023$       & $0.819\pm0.022$        \\
    4 & AdaBoost - SVM - RBF                   & $0.768\pm0.050$        & $0.730\pm0.049$       & $0.796\pm0.031$        & $0.780\pm0.017$       & $0.637\pm0.024$       & $0.808\pm0.014$        \\
    5 & AdaBoost - DT                          & $\pmb{0.908\pm0.017}$  & $0.854\pm0.022$       & $0.895\pm0.014$        & $\pmb{0.901\pm0.007}$ & $\pmb{0.875\pm0.005}$ & $\pmb{0.940\pm0.005}$  \\ \midrule
    6 & AdaBoost - DT - W/O Lungs statistics   & $0.779\pm0.042$        & $0.852\pm0.026$       & $0.877\pm0.024$        & $0.825\pm0.029$       & $0.802\pm0.018$       & $0.886\pm0.019$        \\
    7 & AdaBoost - DT - W/O Opacities statistics    & $0.816\pm0.023$        & $0.832\pm0.028$       & $0.870\pm0.015$        & $0.842\pm0.012$       & $0.819\pm0.010$       & $0.891\pm0.013$        \\
    8 & AdaBoost - DT - W/O Opacities texture       & $0.835\pm0.036$        & $\pmb{0.886\pm0.021}$ & $\pmb{0.910\pm0.011}$  & $0.870\pm0.018$       & $0.843\pm0.016$       & $0.927\pm0.008$        \\
    9 & AdaBoost - DT - W/O Location \& Shape  & $0.856\pm0.018$        & $0.858\pm0.030$       & $0.891\pm0.024$        & $0.873\pm0.007$       & $0.847\pm0.014$       & $0.924\pm0.013$        \\\bottomrule
\end{tabular}}
\caption{Rows $1-5$: Classification performance (mean results) of different classifiers utilizing all depicted features. Rows $6-9$: Ablation study (mean results), obtained while removing each feature group in turn. Results are averaged across 5 folds.}
\label{ReusltsTable}
\end{table*}
\end{center}

\vspace{-10mm}
\section{Experiments and Results}
\label{ExperimentsResults}

\subsection{Classification optimization process \& results}

Data was divided into 5 folds ($80\%\ train,\ 20\%\ test$) for cross validation purposes. Training and testing was repeated throughout the folds to ensure results are indifferent to data division.
Rows $1-5$ in Table \ref{ReusltsTable} present the optimal results, obtained by the different  classifiers. 
The results demonstrate the superiority of the DT based AdaBoost ensemble model, offering a design best suited for both discrete and continues features and dividing the data into class differentiated subgroups. Our AdaBoost - DT based model achieved an optimal point sensitivity of $0.908$ with specificity $0.854$ and a high ROC-AUC of $0.94$. 
Model results show robustness to data division between folds since all metrics demonstrate a small standard deviation.
All metrics indicate a clear dominance of the AdaBoost classifier over the RF. Qualitative examples of our system outputs and classification for 3 different pathological cases are displayed in Figure \ref{keyimages}. 

\subsection{Feature analysis}
\begin{figure}[htb]
 \centering
 \includegraphics[width=8.5cm]{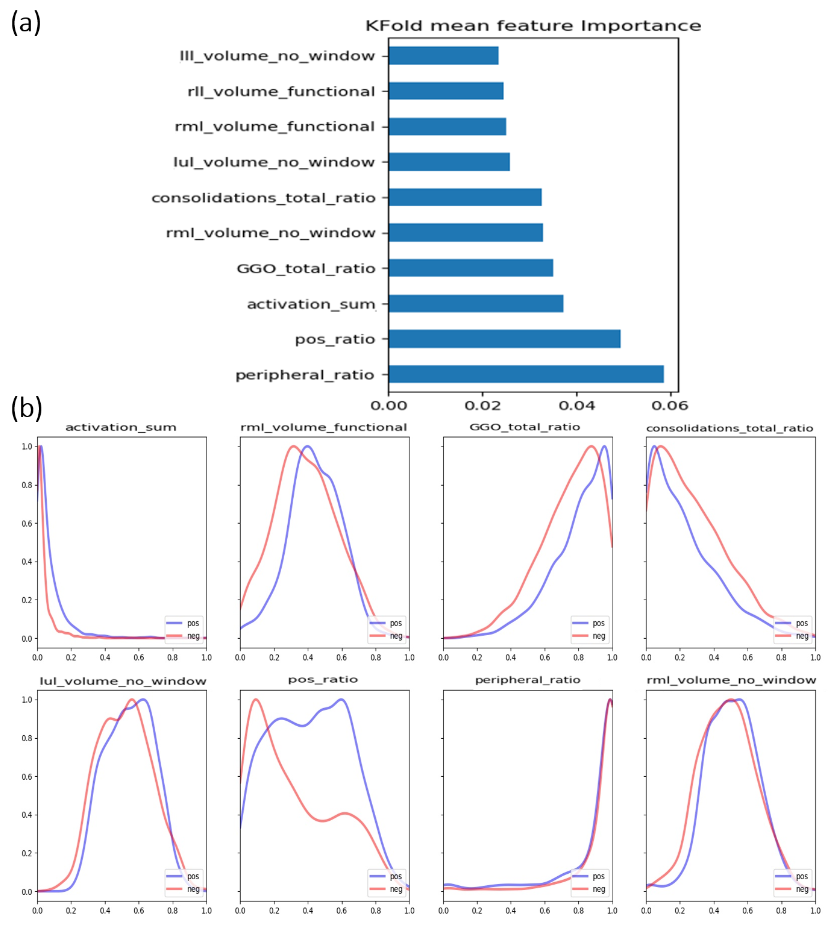}
 \caption{(a) Features importance; (b) KDE analysis on selected features.}
 \label{KDEIMPORTANCE}
\end{figure}

To further assert the importance of the features composing the decision trees weak classifiers in the final ensemble, an analysis of the mean Gini Importance (i.e Mean Decrease in Impurity) per feature across all folds was performed. The top 10 important features are shown in Figure \ref{KDEIMPORTANCE}a. The differences/ similarities in features distributions between COVID-19 positive and negative groups is demonstrated in Figure \ref{KDEIMPORTANCE}b, showing a kernel density estimation (KDE) of several dominant features (normalized to $[0,1]$) obtained using an estimation based on all aforementioned datasets. KDE analysis depicts the probability (y-axis) for each of displayed features to hold a certain value (x-axis) in division to positive and negative patients. It can be noted that some of the dominant features (``pos\_ratio" and ``GGO\_total\_ratio") distributions differ between the classes while others (``peripheral\_ratio" and ``activation\_sum") do not.
This observation, along with these features high importance (high decrease in impurity), as depicted in Figure \ref{KDEIMPORTANCE}a, suggests that a combination of weak classifiers is necessary to enable proper classification. 

\subsection{Feature ablation study}
Among paramount features described in Figure \ref{KDEIMPORTANCE} is a combination of lungs and opacities statistics, opacities texture and location. This indicates the importance of each of the feature groups to the final classification. To assess the quantitative contribution of the groups we conducted an ablation study, removing in turn each of the feature groups from the classification process and measuring the effect on the final ensemble classifier performance. Ablation results (displayed in Table \ref{ReusltsTable}) demonstrate that all ablated groups removal lead to a decrease in almost all measured metrics. We notice that the lungs and opacities statistics feature groups ablations resulted in the most significant metrics decrease.  When removing the lungs statistics (opacities statistics) features sensitivity decreased by $12.9\%$ ($9.2\%$). This large variability in performance indicates the importance of the features to the classification process. An exception is observed when omitting the opacities texture features, which resulted in a slightly increased specificity and precision scores (indicating a lower number of FP), however it goes hand in hand with a drastic drop in sensitivity (i.e increasing number of FN). Since high sensitivity is crucial in the context of emergency disease control this metric was of the highest importance.

\vspace{-0.15cm}
\section{Discussion and Conclusion}
\label{Discussion}
We propose an end-to-end system for COVID-19 chest CT based automated triage. Our model was trained and tested on a large scale database from various sources, containing a diversity of pathological states. The pipeline enables the classification explainability using the extraction of clinical features as well as their quantitative analysis and visualizations.
Results show 90.8\% sensitivity, 85.4\% specificity and accuracy of 87.5\%, thus expanding the previously demonstrated iSARF model \cite{Shi2020} ability for screening between COVID-19 and CAP screening to additional pathological states without compromising the obtained results (90.7\% sensitivity, 83.3\% specificity and accuracy of 87.9\%). The ablation study demonstrated that each of the extracted features holds a unique contribution to the final classification performance. The selected classification approach, excesses the need to normalize feature scales/ units and prioritize between features, being an inherent feature of the weighted ensemble classifier. 
In conclusion, automated chest CT based screening models show great promise in COVID-19 triage, and provide quantitative based analysis of the patient pathological state. In the future, such models could provide diagnosis support as well as contribute to the detection of incidental findings. 

\section{Compliance with Ethical Standards}
All Private datasets cases used in this paper were extracted by querying cloud picture archiving and communication systems (PACS) for cases that contained lung abnormalities. Their usage in this study is in line with the principles of the Declaration of Helsinki.
All data provided was annonimized deleting all patient identifiable information.
Open access datasets usage ethical approval was not required as confirmed by the license attached with the open access data \cite{Zhang2020}.

\bibliographystyle{IEEEbib}
\bibliography{root}

\begin{thebibliography}{10}

\bibitem{Rubin2020}
G.~D. Rubin, C.~J. Ryerson, L.~B. Haramati, N.~Sverzellati, J.~P. Kanne,
  S.~Raoof, and Ann~N. L.,
\newblock ``The role of chest imaging in patient management during the covid-19
  pandemic: a multinational consensus statement from the fleischner society,''
\newblock {\em Radiology}, vol. 296, pp. 172–180, 2020.

\bibitem{Caruso2020}
D.~Caruso, M.~Zerunian, M.~Polici, F.~Pucciarelli, T.~Polidori, C.~Rucci, and
  A.~Laghi,
\newblock ``Chest ct features of covid-19 in rome, italy,''
\newblock {\em Radiology}, vol. 296, pp. 79--85, 2020.

\bibitem{Wang2020}
H.~Wang, R.~Wei, G.~Rao, J.~Zhu, and B.~Song,
\newblock ``Characteristic ct findings distinguishing 2019 novel coronavirus
  disease (covid-19) from influenza pneumonia,''
\newblock {\em European Radiology}, vol. 30, pp. 4910–4917, 2020.

\bibitem{shi2020review}
F.~Shi, J.~Wang, J.~Shi, Z.~Wu, Q.~Wang, Z.~Tang, and D.~Shen,
\newblock ``Review of artificial intelligence techniques in imaging data
  acquisition, segmentation and diagnosis for covid-19,''
\newblock {\em IEEE reviews in biomedical engineering}, 2020.

\bibitem{Ozsahin2020}
I.~Ozsahin, B.~Sekeroglu, M.~S. Musa, M.~T. Mustapha, and D~Uzun~Ozsahin,
\newblock ``Review on diagnosis of covid-19 from chest ct images using
  artificial intelligence,''
\newblock {\em Computational and Mathematical Methods in Medicine}, 2020.

\bibitem{Shi2020}
F.~Shi, L.~Xia, F.~Shan, D.~Wu, Y.~Wei, H.~Yuan, and D.~Shen,
\newblock ``Large-scale screening of covid-19 from community acquired pneumonia
  using infection size-aware classification,''
\newblock {\em arXiv preprint arXiv:2003.09860.}, 2020.

\bibitem{Zhang2020}
K.~Zhang, X.~Liu, J.~Shen, Z.~Li, Y.~Sang, X.~Wu, and L.~Ye,
\newblock ``Clinically applicable ai system for accurate diagnosis,
  quantitative measurements and prognosis of covid-19 pneumonia using computed
  tomography,''
\newblock {\em Cell}, vol. 181, pp. 1423--1433, 2020.

\bibitem{Gozes2020}
O.~Gozes, M.~Frid-Adar, H.~Greenspan, P.~D. Browning, H.~Zhang, W.~Ji, and
  E.~Siegel,
\newblock ``Rapid ai development cycle for the coronavirus (covid-19) pandemic:
  Initial results for automated detection \& patient monitoring using deep
  learning ct image analysis,''
\newblock {\em arXiv preprint arXiv:2003.05037}, 2020.

\bibitem{Gozesmiccai2020}
O.~Gozes, M.~Frid-Adar, N.~Sagie, A.~Kabakovitch, D.~Amran, R.~Amer, and
  H.~Greenspan,
\newblock ``A weakly supervised deep learning framework for covid-19 ct
  detection and analysis,''
\newblock in {\em Proceedings of The Second International Workshop on Thoracic
  Image Analysis}. MICCAI, 2020.

\bibitem{Bernheim2020}
A.~Bernheim, X.~Mei, M.~Huang, Y.~Yang, Z.~A. Fayad, N.~Zhang, and M.~Chung,
\newblock ``Chest ct findings in coronavirus disease-19 (covid-19):
  relationship to duration of infection,''
\newblock {\em Radiology}, vol. 295(3), pp. 685–691, 2020.

\end{thebibliography}

\label{Discussion}
\end{document}